\documentclass[prl, twocolumn, showpacs, english, preprintnumbers, amsmath, amssymb, superscriptaddress, aps]{revtex4-1}
\usepackage{latexsym}
\usepackage{epsfig,psfrag}
\usepackage{dcolumn}
\usepackage{bm}
\usepackage{graphicx}
\usepackage{color}
\usepackage{esint}
\usepackage{babel}
\usepackage{amsfonts}
\usepackage{enumerate}
\begin{document}
\title{Multi-channel Kondo impurity dynamics in a Majorana device} 
\author{A. Altland}
\affiliation{Institut f\"ur Theoretische Physik,  
Universit\"at zu K\"oln, Z\"ulpicher Str. 77, D-50937 K\"oln, Germany}
 \author{B. B{\'e}ri}
\affiliation{
School of Physics and Astronomy, University of Birmingham, Edgbaston, 
Birmingham B15 2TT, UK}
\author{R. Egger}
\affiliation{ Institut f\"ur Theoretische Physik, Heinrich-Heine-Universit\"at, D-40225 D\"usseldorf, Germany}
\author{A.M. Tsvelik}
\affiliation{Department of Condensed Matter Physics and Materials Science, 
Brookhaven National Laboratory, Upton, NY 11973-5000, USA} 
\date{\today}

\begin{abstract}
We study the multi-channel Kondo impurity dynamics 
realized in a mesoscopic superconducting island connected to metallic leads.  
The effective ``impurity spin'' is non-locally realized by Majorana 
bound states and strongly coupled to lead electrons by 
non-Fermi liquid correlations.  
We explore the spin dynamics and its observable ramifications near the 
low-temperature fixed point. The topological protection of the system 
raises the perspective to observe multi-channel Kondo impurity dynamics
in experimentally realistic environments. 
\end{abstract}

\pacs{71.10.Pm, 73.23.-b, 74.50.+r} 

\maketitle

\textit{Introduction.---}The coupling of a local quantum degree of 
freedom (``impurity'') to an ideal Fermi gas can generate strong 
correlations which ultimately may push the system outside the realm
of the Fermi liquid. The best known realization of this phenomenon is the
multi-channel Kondo effect in the overscreened regime $M>2S$, 
where the impurity is a local spin-$S$ degree of
freedom and the Fermi gas is realized 
as a set of $M$ one-dimensional conduction channels. 
The multi-channel Kondo effect has been the subject of a huge body 
of theoretical studies, and the non-perturbative mechanisms behind 
the formation of its non-Fermi liquid fixed points are by now well understood \cite{blandin,wiegmann,andrei,tsvelik,AL,gogolinbook,hewson}. At the same
time, the experimental realization of this seemingly rather basic system is met with severe difficulties \cite{gordon}, since 
anisotropies in the couplings between impurity and different channels are 
relevant perturbations \cite{blandin}. As a consequence,  
a delicate fine tuning of coupling constants is required, a condition few if any realizations of the system are able to meet. For the same reason, the highly entangled effective degrees of freedom predicted to form at strong coupling 
\cite{wiegmann,andrei,tsvelik,AL} have so far 
remained beyond experimental access. 

\begin{figure}
\centering
\includegraphics[width=8cm]{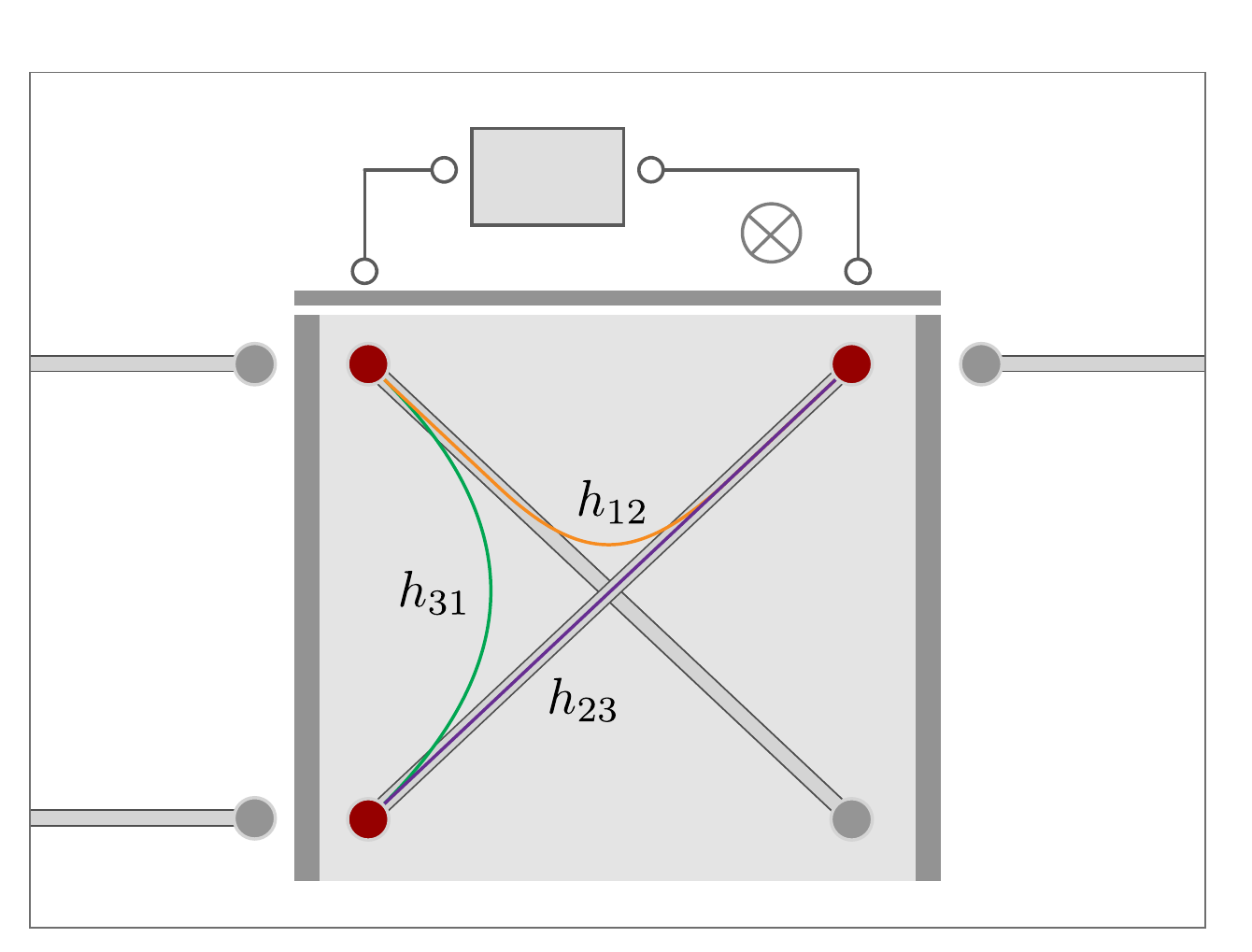}
\caption{\label{fig1} 
(Color online) Schematic setup leading to the topological Kondo effect. A
floating superconducting island (center square) with charging energy $E_c$ supports a helical crossed nanowire \cite{leo}; alternative 
realizations using several nanowires are also possible. At the terminal points of the wires, Majorana fermions $\gamma_j$ (the circles) are present \cite{mbsrev1,mbsrev2,mbsrev3},
$M$ of which are coupled to external leads (here $M=3$). 
Direct tunnel couplings $h_{jk}$ between the Majoranas act like a
Zeeman field on this ``Majorana impurity spin''. The upper part 
illustrates the optional coupling to a single-electron box via
flux-tunable tunnel amplitudes. This provides a 
way~\cite{flensberg} to read out the effective ``magnetization''
 $\sim i \left\langle \gamma_j
\gamma_k\right\rangle$. }
\end{figure}

The recently proposed ``topological'' Kondo effect \cite{beri1} promises a 
rather more robust realization of non-Fermi liquid correlations. 
In this system, schematically indicated in Fig.~\ref{fig1}, the
``impurity'' is formed by the $M_{\rm tot}$ Majorana end states of 
spin-orbit coupled quantum wires in proximity to a finite piece 
of $s$-wave superconducting material \cite{mbsrev1,mbsrev2,mbsrev3} 
--- a setup realizable by current device technology
 \cite{exp1,exp2,exp3,exp4,exp5,exp6}. The mutual coupling $h_{ij}$
 between Majorana bound states $i,j=1,\dots,M_{\rm tot}$ 
is often significant and can be tuned by external gates; 
in the spin analogy, it plays the role of an effective Zeeman field. 
Tunnel coupling $M\le M_{\rm tot}$ Majoranas to normal leads, see 
Fig.~\ref{fig1}, generates an effective 
Kondo setup, where the ``reality'' of the compound Majorana states 
implies that $\mathrm{SO}(M)$ rather than the more
 conventional $\mathrm{SU}(2)$ (but see Refs.~\cite{avishai1,avishai2}) 
plays the role of the symmetry group.
Importantly, the non-Fermi liquid Kondo fixed point of such a
device is self-stabilizing: Regardless of disparities in the lead-to-Majorana 
tunnel couplings or other sources of channel anisotropy,
it will be approached at low temperatures.
Signatures of this flow in, e.g., the power-law scaling of 
conductance coefficients  have been the subject of 
Refs.~\cite{beri1,altland1,beri2,altland2}. 

However, arguably the most striking manifestation of quantum criticality
in the multi-channel Kondo effect is the formation of a massively 
entangled effective degree of freedom
governing the system at strong coupling. In this Letter, we argue that
mesoscopic Majorana devices offer, for the first time, a
perspective to probe and manipulate such type of quantum degrees of 
freedom within a challenging yet realistic experimental setup. 
The readout observables in this context are various transport 
coefficients and ``spin'' expectation values,   
 while the ``knobs'' to manipulate the non-Fermi liquid 
impurity are the Zeeman field coefficients mentioned above. 
Our main findings are summarized as follows: 
(i)  We show that the Zeeman field
strongly affects the conductance coefficients, $G_{jk}$, as well as the
``magnetization'', i.e., the expectation value of the Majorana spin components. 
On intermediate temperature scales, both observables can be accessed by
perturbation theory around the fixed point. 
 We also propose a scheme to experimentally probe the
 magnetization in terms of a protocol involving a joint tuning of the Zeeman field and the tunnel couplings to the leads. (ii) We show that the Majorana spin has nonvanishing $M$-point ($M/2$-point) correlation functions for odd (even) $M$, and how these imply the existence of nonlinear susceptibilities and frequency mixing. In the conventional two-channel SU(2) Kondo model, such effects are absent. (iii) For very low temperatures, the Zeeman field destabilizes the Kondo fixed point. Specifically, a field with just one component drives a crossover between two
Kondo fixed points with $M\to M-2$, which will manifest itself in a definite change of, e.g., the temperature dependence of transport coefficients.

Generally speaking,  the macroscopic realization of our ``impurity'' 
in terms of long-range entangled Majoranas should make the system 
 more accessible than a ``real'' spin or other
microscopic few-level systems. It is also worth
pointing out that all phenomena listed above essentially rely on the 
``Zeeman'' coefficients $h_{ij}$, i.e., on couplings generally considered 
obstructive to the observation of Majoranas. 
In the present context, the dependence of observables on these 
quantities is a defining element of the theory, and it stands to reason that
 the observation of any of the effects (i)-(iii) would
 provide compelling evidence for Majorana fermions.

\textit{Model.---} We consider a setup as shown
schematically in Fig.~\ref{fig1}. The set of $M$ Majorana fermions
tunnel connected to leads is described by operators
$\gamma_j=\gamma_j^\dagger$ subject to the Clifford algebra $
\{\gamma_j,\gamma_k\} = 2\delta_{jk}$ \cite{mbsrev1,mbsrev2,mbsrev3}. 
The $\{ \gamma_j \}$ compose a spinor representation of the SO($M$) group, and
the $M(M-1)/2$ different products $i\gamma_j\gamma_k$ define the components of
the Majorana spin.  The Hamiltonian describing the system at energy
scales below the island charging energy, $E_c$, is 
\cite{beri1,altland1,beri2,altland2}
\begin{eqnarray} \label{model}
H &=& -i\sum_{j=1}^M \int_{-\infty}^\infty dx \ \psi_j^\dagger(x) 
\partial_x \psi_j^{}(x) \\ 
\nonumber &+& \sum_{j\neq k} \lambda_{jk}
\gamma_{j} \gamma_{k} \psi_{k}^{\dagger}(0) \psi^{}_{j}(0)
+ i\sum_{j\neq k} h_{jk}\gamma_j\gamma_k,
\end{eqnarray}
where $\psi_{j}(x)$ is an effectively spinless right-moving fermion field
describing the $j$th lead; unfolding from $x<0$ to the full line is
understood, with $x=0$ at the tunnel contact. (We set the  Fermi velocity
$v=1$ and use units with $\hbar=k_B=1$.) 
The symmetric matrix of ``exchange couplings'' is given by
$\lambda_{jk}\approx t_j t_k/E_c >0$, with lead-Majorana tunnel couplings $t_j$,
 while the direct couplings $h_{jk}=-h_{kj}$ 
between Majoranas act like Zeeman fields. 
Concerning the remaining $M_\text{tot}-M$ Majoranas on the island
which are not coupled to  leads, we assume that these have no 
direct tunnel couplings with the $\gamma_j$ \cite{footmidgap}.  For $h_{jk}=0$
and on energy scales below the Kondo temperature
\begin{equation}\label{tk}
T_K\simeq E_c \exp\left(- \frac{\pi}{(M-2) \bar \lambda}\right),
\end{equation}
with average exchange coupling $\bar \lambda$, the model (\ref{model}) scales
to the topological Kondo fixed point \cite{beri1}. 
In contrast to conventional multi-channel Kondo systems \cite{blandin,wiegmann,andrei,tsvelik,AL,gogolinbook,hewson,fiete}, 
anisotropy in the $\lambda_{jk}$ is an irrelevant perturbation
near the fixed point, which in turn corresponds to
an SO$_2(M)$ 
Wess-Zumino-Novikov-Witten boundary conformal field theory 
(BCFT) \cite{AL,CFT}. While the ensuing physics can be discussed 
within the framework of the Affleck-Ludwig \cite{AL} BCFT approach 
(adapted to the SO$_2(M)$ case), it also admits a more direct 
bosonization description.

\textit{Abelian bosonization.---}In terms of bosonization, 
the lead Hamiltonian in Eq.~(\ref{model}) is represented as
$H_\text{lead}= \frac{1}{8\pi} \sum_j \int dx [ 
(\partial_{x}\theta_j)^{2}+(\partial_{x}\varphi_j)^{2} ],$
where $\theta_j$ and $\varphi_j$ are dual bosonic fields
\cite{gogolinbook,delft} defined for $x< 0$, 
with boundary condition $\varphi_j(0)=(\partial_x\theta_j)(0)=0$ 
at the tunnel contact.  In addition to the $\gamma_j$, the exchange coupling
term in Eq.~(\ref{model}) involves the lead electron 
operators at $x=0$, which are represented by 
$\psi_j(0)=  \frac{i}{\sqrt{a}}\Gamma_j
e^{i\theta_j(0)/2}$ \cite{nayak,oshikawa} with the 
short-distance length $a$.  The Klein factors $\Gamma_j=\Gamma_j^\dagger$ 
establish anticommutation relations between 
electrons on different leads \cite{delft}, 
$\{\Gamma_j,\Gamma_k\}=2\delta_{jk}$ and $\{\Gamma_j,\gamma_k \}=0,$
and can be represented as auxiliary 
Majorana fermions \cite{beri2,altland1,altland2}. 
Redefining $\lambda_{jk}\to a\lambda_{jk}$,  the exchange term then becomes 
$H_K=\sum_{j\neq k} \lambda_{jk} \gamma_j\Gamma_j
\gamma_k\Gamma_k e^{-i[\theta_{k}(0)-\theta_{j}(0)]/2}$.

Combining physical Majoranas and Klein factors
to the ``hybrid'' fermion operators $d_j=(\gamma_j+i\Gamma_j)/2$,
both types of Majorana fermions enter only through the products 
$p_jp_k$ of the parities of the hybrid fermions shared between them, 
$p_j=i\gamma_j\Gamma_j= 2 d_j^\dagger d_j^{}-1 = \pm 1.$
As parity products commute, they can be simultaneously diagonalized, 
$p_j p_k= \pm 1$.  At first sight, the $p_j$ themselves seem to generate 
good quantum numbers. However, the total fermion parity also
has to be conserved, $P_\text{tot}\sim
\prod_{j=1}^{M_\text{tot}} \gamma_j=\pm 1$.
This parity constraint holds when no above-gap quasiparticles
are accessible, and reflects the fixed total electron number
on the island within the topological Kondo regime.
Since $\{p_j,P_\text{tot}\}=0$, the parity constraint is violated by
individual $p_j$ operators.  Similar constraints on compound Majorana systems 
have been discussed recently~\cite{fu2010,njpe,wilczek}, and must
be taken into account on top of the Clifford algebra.
Here we have $M-1$ independent conserved parity products, e.g., $p_j p_M=\pm 1$,
such that we arrive at a 
purely bosonic problem for given collection $\{ p_jp_k\}$.

\textit{Kondo fixed point.---}For $h_{jk}=0$, the renormalization 
group (RG) flow of the $\lambda_{jk}$ proceeds towards an 
isotropic strong-coupling fixed point.  
The effect of this becomes transparent after an orthogonal rotation of 
the boson fields, $\boldsymbol{\theta}=(\theta_1,\ldots,
\theta_M)$ and $\boldsymbol{\varphi}=(\varphi_1,\ldots,\varphi_M)$.
Using the unit vector $\mathbf{v}_0=\frac{1}{\sqrt M}(1,\ldots,1)$,
we decompose them to $\theta_0=\mathbf{v}_0\cdot \boldsymbol{\theta}$ and 
$\varphi_0=\mathbf{v}_0\cdot \boldsymbol{\varphi}$, with  
the remaining $M-1$ components $\tilde \theta_j$ and $\tilde \varphi_j$, 
respectively, along the directions orthogonal to $\mathbf{v}_0$.
This rotation decouples the $(\theta_0,\varphi_0)$ sector, and 
the exchange term becomes
\begin{equation} \label{eq:qbmform}
H_K=-\sum_{j\neq k} \lambda_{jk} p_j p_k \exp\left( \frac{i}{2} 
( \mathbf{w}_k- \mathbf{w}_j ) \cdot \boldsymbol{\tilde\theta}(0) \right).
\end{equation}
The $M$ vectors ${\bf w}_j$ are of dimension $M-1$,  with
$\mathbf{w}_{j}\cdot\mathbf{w}_{l}=\delta_{jl}-1/M$,
and span the field space orthogonal to the zero modes.
For each set $\{ p_j p_k \}$, Eq.~(\ref{eq:qbmform}) defines a 
boundary potential with minima forming a hyper-triangular lattice
\cite{nayak,oshikawa,YiKane,YiQBM}.
Near the $\lambda_{jk}\to \infty$ fixed point, $\boldsymbol{\tilde\theta}(0)$ 
tends to be pinned to one of these $\{ p_j p_k\}$-dependent minima. 
The weak-coupling Neumann boundary conditions
of $\boldsymbol{\theta}$ are thus dynamically replaced by Dirichlet conditions
for $\boldsymbol{\tilde\theta}$ near the Kondo limit.  
For $h_{jk}=0$, the leading perturbations are due to
operators preserving $\{ p_jp_k \}$ while tunneling 
$\boldsymbol{\tilde\theta}(0)$ between adjacent
minima \cite{beri2,altland1}.  These perturbations are 
RG irrelevant, of scaling dimension $\Delta_{\rm irr}=1+\frac{M-2}{M} >1$, 
consistent with a stable fixed point. $\Delta_{\rm irr}$ also coincides 
with that
of the first descendant of the adjoint primary, which we identify as 
the leading perturbation of the SO$_2(M)$ BCFT.  Near this fixed point, 
the high-energy cutoff scale of the theory is then set by the 
Kondo temperature $T_K$ in Eq.~(\ref{tk}).

\textit{Majorana spin at strong coupling.---}The  Zeeman term  
generates RG relevant perturbations which destabilize the 
SO($M$) Kondo fixed point.  In terms of symmetries, this is because the Zeeman field breaks an emergent time-reversal invariance  
of Eq.~(\ref{model}).  In terms of bosonization, the reason is  
that  $\gamma_j\gamma_k$ does not commute with all possible products
$p_m p_n$.  This implies that additional tunneling processes 
can take place, where $\boldsymbol{\tilde\theta}(0)$  connects minima  
belonging to different $\{ p_j p_k \}$ sectors.  
For short enough tunneling ``distance'', 
such a process becomes RG relevant. 
 The corresponding scaling operator ${\cal S}_{jk}$ conjugate to $h_{jk}$ can be inferred from symmetry arguments. In particular, ${\cal S}_{jk}$ should 
(i) conserve all $p_m p_n$ products commuting with $\gamma_j\gamma_k$, 
(ii) respect the $\boldsymbol{\tilde\theta}$ Dirichlet conditions, 
(iii) should have the same SO($M$) rotational properties as the 
Zeeman perturbation, (iv) commute with $P_{\rm tot}$, (v) conserve charge, and  
(vi) be a local operator acting at the ``impurity'' position.
Conditions (i) to (vi) determine \cite{SM} the bosonized 
representation of the Majorana spin near the Kondo fixed point:
${\cal S}_{jk}\sim {\cal S}_{jk}^{(+)}+{\cal S}_{jk}^{(-)}$ consists of
the ``bare'' operator dressed by bosonic phase factors,
\begin{equation} \label{eq:Opm}
{\cal S}_{jk}^{(\pm)} = i\gamma_{j}\gamma_{k}
\cos\left(\frac12 (\mathbf{w}_{j}\pm \mathbf{w}_k) \cdot
\boldsymbol{\tilde\varphi}(0)\right).
\end{equation}
Depending on the sign, the operators in Eq.~(\ref{eq:Opm}) have dimension 
$\Delta_+=1-\frac{2}{M}$ (relevant) or $\Delta_-=1$ (marginal),
identical to those of the adjoint primary and the descendant of the 
identity, respectively. We have thus identified these operators as 
the leading time-reversal symmetry breaking perturbations of the 
SO$_2(M)$ BCFT.  For small $h_{jk}$, it suffices to keep only
 the RG relevant operator in the Zeeman term.  
This still represents a weak perturbation  around the 
Kondo fixed point on intermediate energy scales, 
$T_h\ll E\ll T_K$, with $T_K$ in Eq.~(\ref{tk}).  
Dimensional scaling yields the Zeeman scale
\begin{equation}\label{th}
T_h = T_K (\bar h/T_K)^{M/2},\quad \bar h={\rm max}\left|h_{jk} \right|.
\end{equation}
On energy scales below $T_h$, the Zeeman field drives the 
system away from the Kondo fixed point and nonperturbative methods are needed. 

\begin{figure}
\centering
\includegraphics[width=8cm]{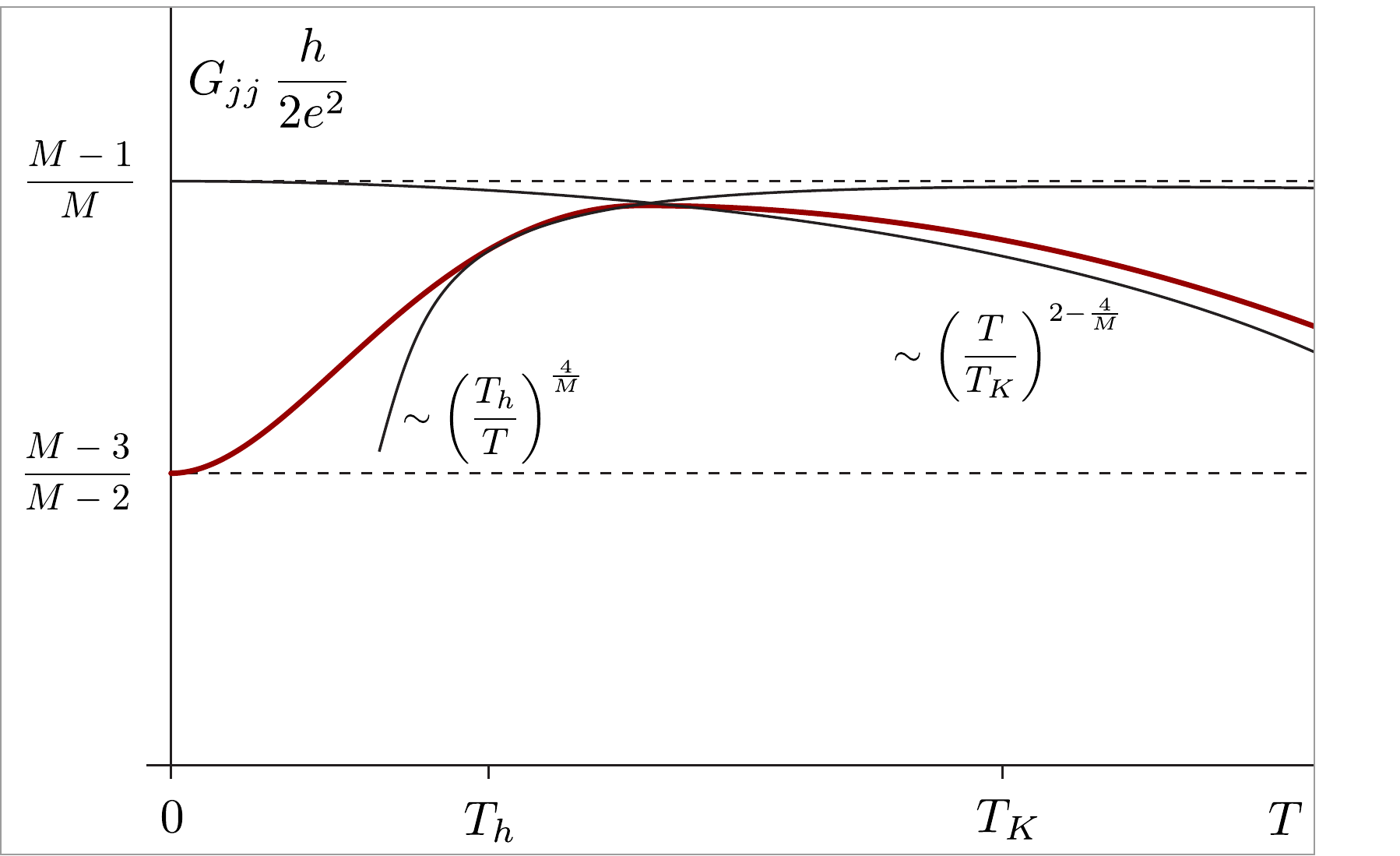}
\caption{\label{fig2} 
(Color online) Schematic sketch of the temperature ($T$) dependence of the 
linear conductance coefficient $G_{jj}$, with $j\ne 1,2$,
for $M>4$ and a Zeeman field with $h_{12}\ne 0$.  For $T\ll T_h$,
the conductance approaches $\frac{2e^2}{h}\frac{M-3}{M-2}$,
as appropriate for the SO$(M-2)$ fixed point. For $T_h\ll T\ll T_K$,
on the other hand, it is the SO($M$) fixed point that governs the 
conductance, with
$G_{jj}\simeq \frac{2e^2}{h}\left(\frac{M-1}{M} - c_1(T_h/T)^{4/M}
-c_2(T/T_K)^{2-4/M}\right)$, where $c_{1,2}$ is of order unity.
}
\end{figure}

\textit{Charge transport.---}The currents $I_j$ flowing through the $j$th 
lead and the respective chemical potentials $\mu_j$ define the 
conductance tensor $G_{jk}=-e \frac{\partial I_j}{\partial \mu_k}$.
Using the above bosonization approach, perturbation theory in the Zeeman
field yields the linear conductances for $T_h\ll T\ll T_K$,
\begin{eqnarray} \label{cond}
\frac{G_{jk}}{2e^2/h} & = &   Q_{jk} - \frac{\sin(2\pi/M)}{M} 
\left(\frac{T_h}{T}\right)^{4/M} 
\sum_{l\ne m} \frac{h_{lm}^2}{\bar h^2}
\\ \nonumber  &\times & (Q_{jl}+Q_{jm}) (Q_{lk}+Q_{mk}) + O(h_{jk}^4),
\end{eqnarray}
with $Q_{jk}= \mathbf{w}_{j}\cdot \mathbf{w}_{k}= \delta_{jk}-1/M$.
It is instructive to analyze the diagonal term, $G_{jj}$, for just 
one non-zero Zeeman component, say $h_{12}$.
Taking $j\ne 1,2$, such that $\gamma_j$ is not Zeeman-coupled 
to other Majoranas, Eq.~(\ref{cond}) predicts a \textit{reduction}
with respect to the $h_{12}=0$ result, 
$G_{jj}=\frac{2e^2}{h}\frac{M-1}{M}$.  
 This reduction can be intuitively understood
by noting that for $T\ll T_h$, the Zeeman field effectively 
removes the two Majoranas $\gamma_1$ and $\gamma_2$ from the low-energy
sector, and thereby drives the system to a new fixed point with
 $M\to M-2$.  For $T\ll T_h$ (and $M>4$), the conductance should 
therefore 
approach the \textit{smaller} value 
$G_{jj}= \frac{2e^2}{h} \frac{M-3}{M-2}$. 
This scenario has also been found by exact Bethe ansatz 
calculations \cite{nextpaper}.  The temperature dependence of the above
conductance coefficient is illustrated in Fig.~\ref{fig2}.

\textit{Multi-point correlations.---}Let us now address
the correlation functions of the Majorana spin components
at the Kondo fixed point. 
Two-point correlations are always diagonal and given by
$\langle {\cal T}_\tau {\cal S}^{(+)}_{jk}(\tau) 
{\cal S}^{(+)}_{jk}(0) \rangle \simeq
 |T_K \tau|^{-2+4/M}$, where ${\cal T}_\tau$ denotes imaginary time ordering.  In fact, this correlator directly implies 
Eq.~(\ref{cond}) for the conductance tensor.
However, the Majorana spin also exhibits remarkable multi-point correlations. For clarity, we first consider the case $M=3$, 
where the three Majorana spin components define
a vector with $S_j=\sum_{kl} \varepsilon_{jkl} {\cal S}_{kl}^{(+)}$.
Within a Coulomb gas interpretation for the bosonized expression 
of the $S_j$, see Eq.~(\ref{eq:Opm}), ``neutral'' phase combinations 
are required for the existence of multi-point correlators \cite{footfree}. This leads to the three-point correlator,
\begin{equation}\label{3pt}
\left\langle {\cal T}_\tau \left[ S_{j}(\tau_1)
S_{k} (\tau_2) S_l(\tau_3) \right] \right\rangle \simeq
\frac{\varepsilon_{jkl}}{T_K (\tau_{12}\tau_{13}\tau_{23})^{1/3}},
\end{equation}
where $\tau_{jk}=\tau_j-\tau_k$ with all $T_K|\tau_{jk}|\gg 1$,
and  $\tau^{1/3}={\rm sgn}(\tau) |\tau|^{1/3}$.
Equation~(\ref{3pt}) is consistent with the SO(3) group 
structure and BCFT fusion rules \cite{AL,CFT}, and has observable
consequences in the ``spin response'' to the Zeeman 
field vector $\boldsymbol{h}$ with $h_j= \sum_{kl} \varepsilon_{jkl} h_{kl}$. In fact, perturbation theory in
$\boldsymbol{h}$ entails from Eq.~(\ref{3pt}) the effective action contribution
\begin{equation}\label{seff}
S_{\rm eff} \sim \int dt_1 dt_2 dt_3
\frac{\boldsymbol{h}(t_1) \cdot [ \boldsymbol{h}(t_2)
\times \boldsymbol{h}(t_3)]}{T_K (t_{12}t_{13}t_{23})^{1/3}},
\end{equation}
where we switch to real time, $\tau\to it$, and allow for a 
time-dependent Zeeman field. 
Taking $\boldsymbol{h}(t)=( h_1\cos[\omega_1 t], 
h_2 \cos[\omega_2 t],0)$, the action (\ref{seff}) implies 
nonlinear frequency mixing, i.e., a finite magnetization 
$\langle S_3(t)\rangle$ that oscillates in time with 
frequencies $\omega_1\pm \omega_2$.
Similarly, for $\boldsymbol{h}(t)=( 0,0, h_3 \cos[\omega t])$ 
with $\omega\gg T_h$, 
the above action predicts that an oscillatory ``transverse'' 
spin correlation function is generated ($t_1>t_2$),
\begin{eqnarray}\label{dynHall} 
\langle S_1(t_1)S_2(t_2)\rangle &\sim&  
h_3 \cos[\omega(t_1+t_2)] F[\omega (t_{1}-t_2)],\\ \nonumber 
F(y)&=& (y/2)^{-1/6} [Y_{-1/6} (y)+ J_{1/6}(y)],
\end{eqnarray}
with the Bessel functions $Y_\nu$ and $J_\nu$.
Since $F(y\gg 1)\simeq y^{-2/3}\cos(y-\pi/3)$,  the
envelope of the oscillatory correlations in Eq.~(\ref{dynHall})
has the slow algebraic long-time tail $\sim (t_1-t_2)^{-2/3}$. 
Finally, we note that similar multi-point correlations appear also for $M>3$.  For odd $M$, the Coulomb gas neutrality condition allows for
$M$-point correlations, while the $M/2$-point correlator may 
survive for even $M$ \cite{SM}.

\textit{Experiment.---}How can the above predictions  be tested experimentally? Our predictions for the conductance should be readily observable in charge transport once the Kondo regime $T\ll T_K$ has been reached; similar experiments (but away from the
Kondo regime) have been carried out previously \cite{exp1,exp3,exp5}. While the observation of the magnetization components is less
straightforward, the presence of definite multi-point correlations~\eqref{3pt} makes them  particularly interesting observables. For a single nanowire with $M=2$ Majoranas, several readout
schemes have been proposed before in the context of topological quantum
computing \cite{mbsrev1,mbsrev2,mbsrev3,burnell}, by employing, e.g., a nearby 
quantum dot \cite{flensberg} or a flux qubit \cite{hassler}. Building on these ideas, we here propose to probe the 
``magnetization'', 
$\sim i \left\langle \gamma_j\gamma_k\right\rangle$,
 via the occupation of the qubit state associated to the non-local fermion $c=(\gamma_j + i\gamma_k)/2$. The readout of this state might proceed in three steps: (i) Switch off all Zeeman couplings and decouple all leads,
e.g., by ramping up gates indicated by vertical bars in Fig.~\ref{fig1},
and reduce the charging gap of the island, e.g., by the gate voltage. 
(ii) Tunnel couple the end states $j,k$ to a single-electron box. 
(iii) The occupation of the $c$ fermion state may now be
 probed~\cite{flensberg}  by detecting the charge state of the 
single-electron box as function of its charging energy and of 
a magnetic flux threading the system, see Fig.~\ref{fig1}.
 
To conclude, we have studied the dynamics of the effective quantum impurity
spin formed by the spatially separated Majorana fermions in a 
topological Kondo device as shown in Fig.~\ref{fig1}.  
This highly unconventional spin exhibits rich and observable 
dynamics characterized by nonvanishing multi-point correlations and 
nonperturbative crossovers between different non-Fermi liquid Kondo 
fixed points.  We hope that the effects predicted here can soon be 
observed  experimentally. 

We thank E. Eriksson, A.A. Nersesyan, V. Kravtsov, and A. Zazunov 
for valuable discussions, and acknowledge financial support by 
the SFB TR12 and the SPP 1666 of the DFG, a Royal Society URF, and the
DOE under Contract No.~DE-AC02-98CH10886.

\newpage
\clearpage
\setlength{\topmargin}{-1.35in}
\setlength{\oddsidemargin}{-1.15in}
\includegraphics[page={1},width=1.15\textwidth]{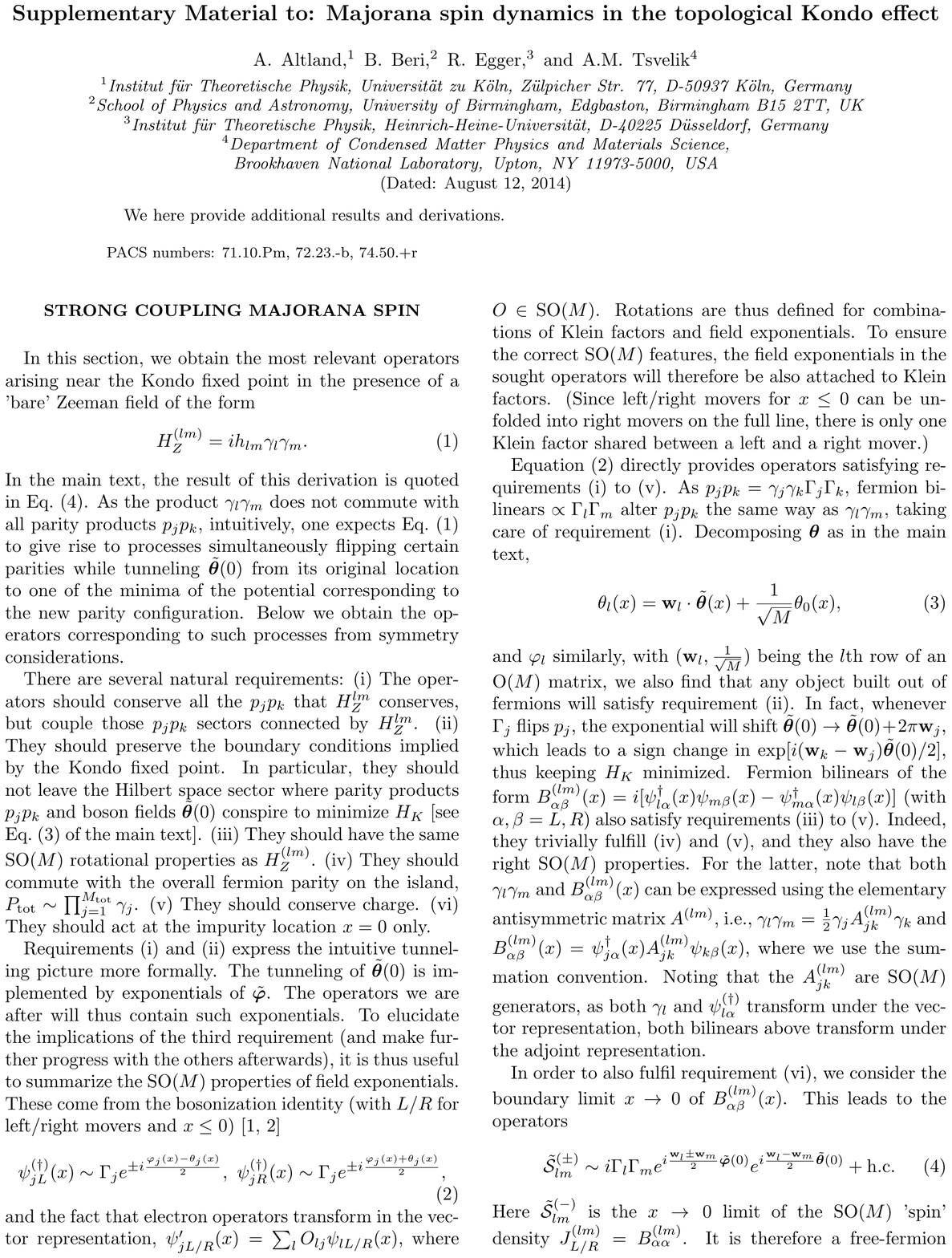}
\clearpage
\setlength{\oddsidemargin}{-1.15in}
\includegraphics[page={2},width=1.15\textwidth]{SM.pdf}
\clearpage
\setlength{\oddsidemargin}{-1.15in}
\includegraphics[page={3},width=1.15\textwidth]{SM.pdf}
 \clearpage

\begin{thebibliography}{99}
\bibitem{footmidgap} 
We assume that 
 the energies of quasiparticle states besides the Majorana end 
states to be higher than the Kondo temperature. On general grounds, 
one expects the quasiparticle energies to be comparable to the 
proximity gap amplitude which, likewise, represents a high energy 
scale in our approach.
\bibitem{footfree} Note that a ``bare'' SO(3) Majorana spin also has
three-point correlations. However, these do not
depend on the $|\tau_{jk}|$ and do not cause the effects discussed here.
\end{thebibliography}
\end{document}